\documentclass[12pt]{article}

\usepackage[reqno]{amsmath}

\usepackage{amssymb,amsthm,graphicx,verbatim,url,verbatim,longtable,vmargin,accents}
\usepackage{float}
\usepackage{fancyvrb}
\allowdisplaybreaks
\usepackage{times,subfig,dcolumn,booktabs,setspace,soul,latexsym,wasysym,titling}
\usepackage[export]{adjustbox}
\usepackage[T1]{fontenc}\usepackage[encapsulated]{CJK}
\usepackage[utf8]{inputenc}
\usepackage{hyperref}
\usepackage{hyperref}

\usepackage[natbib=true,backend=biber,style=authoryear,citestyle=authoryear,ibidtracker=false,maxcitenames=2,uniquename=false,uniquelist=false]{biblatex}

\makeatletter
\def\@biblabel#1{}
\makeatother

\usepackage[all]{xy}
\newcolumntype{.}{D{.}{.}{-1}}\newcolumntype{d}[1]{D{.}{.}{#1}}
\usepackage[dvipsnames,usenames]{color}\definecolor{spot}{rgb}{0.6,0,0}

\usepackage{atbegshi}

\usepackage[titletoc,title]{appendix}

\newcommand{\TheoremName}{Observation}

\newcommand{\E}{\mathbb{E}}
\newcommand{\bX}{\mathbf{X}}
\newcommand{\bD}{\mathbf{D}}

\newcommand{\bW}{\mathbf{W}}
\newcommand{\bbeta}{\boldsymbol{\beta}}

\newcommand{\bY}{\mathbf{Y}}

\newcommand{\bT}{\mathbf{T}}

\newcommand{\btau}{\boldsymbol{\tau}}

\newcommand{\blind}{0} 
\newcommand{\titl}{Degrees of Randomness in Rerandomization Procedures}

\if0\blind
 \title{\titl}
 \title{\titl}
 \author{Connor T.\ Jerzak\thanks{
Assistant Professor, Government Department, The University of Texas at Austin. Email:
\href{mailto:connor.jerzak@austin.utexas.edu}{connor.jerzak@austin.utexas.edu} URL:
\href{https://www.connorjerzak.com}{ConnorJerzak.com} ORCID: 0000-0001-9858-2050}\and 
Rebecca Goldstein\thanks{Assistant Professor of Law, Jurisprudence and Social Policy Program, UC Berkeley School of Law, 2240 Piedmont Avenue, Berkeley CA 94720. Email:
\href{mailto:rgoldstein@berkeley.edu}{rgoldstein@berkeley.edu} URL:
\href{https://rebeccasgoldstein.com/}{RebeccasGoldstein.com} ORCID: 0000-0002-9944-8440}
} 

\fi\if0\blind
\title{\titl}
\fi

\begin{document}
\newpage 
\maketitle
\begin{abstract}
\noindent Randomized controlled experiments are susceptible to imbalance on covariates predictive of the outcome. Rerandomization and deterministic treatment assignment are two proposed solutions. This paper explores the relationship between rerandomization and deterministic assignment, showing how deterministic assignment is an extreme case of rerandomization. The paper argues that in small experiments, both fully randomized and fully deterministic assignment have limitations. Instead, the researcher should consider setting the rerandomization acceptance probability based on an analysis of covariates and assumptions about the data structure to achieve an optimal alignment between randomness and balance. This allows  for the calculation of minimum $p$-values along with valid permutation tests and fiducial intervals. The paper also introduces tools, including a new, open-source \textbf{\textsf{R}} package named \texttt{fastrerandomize}, to implement rerandomization and explore options for optimal rerandomization acceptance thresholds.
\\ {\bf Keywords:} Rerandomization; 
Design of experiments; 
Design-based inference;
Optimally balanced randomization
\end{abstract}


\setcounter{page}{1}
\section{Introduction}
Although randomized controlled experiments are now a cornerstone of causal inference in the medical and social sciences, researchers have known since at least \citet{Gosset38} that in finite samples, random assignment will often result in imbalance between the treated and control groups on covariates predictive of the outcome. This imbalance will generate conditional bias on estimates of the average treatment effect while adding noise to treatment effect estimates.

In a 2012 paper, Morgan and Rubin propose a solution: rerandomization \citep{MorganRubin2012}. They argue that if experimenters provide a quantitative definition of imbalance in advance, they can safely discard randomizations that are too imbalanced and obtain more precise estimates of treatment effects. After rerandomization, researchers adjust exact significance tests and fiducial intervals by excluding randomizations that would not have been accepted according to the initial acceptance threshold. Researchers can also calculate fiducial intervals by exploiting the duality between intervals and tests. Rerandomization has already gained traction as a best practice for field researchers: it is included as a recommended practice for minimizing covariate imbalance in the Handbook of Field Experiments \citep{DufBan17}. 

At the same time, \citet{Kasy2016} revived a decades-old debate between Gossett (writing as ``Student'') and Fisher by arguing that ``experimenters might not want to randomize in general'' because ``[t]he treatment assignment that minimizes conditional expected loss is in general unique if there are continuous covariates, so that a deterministic assignment strictly dominates all randomized assignments.'' A similar argument is made in \citet{Sav59,BerJohKal15,Kal18} and proven in a theoretical framework by \citet{BanChaMon17}, who show that Bayesians and near-Bayesians should not randomize. 

In this paper, we begin by formalizing the idea that these two solutions to the problem of covariate imbalance in randomized controlled trials -- rerandomization (the Morgan and Rubin approach) and deterministic choice of an optimally balanced treatment assignment vector (the Kasy approach)---are closely related but very much in conflict.\footnote{\citet{Kasy2016} discusses the relationship between the two ideas in concept, writing that ``I very much agree with this [Morgan and Rubin's] argument; one way to think of the present paper is that it provides a formal foundation for this argument and takes it to its logical conclusion.''} In particular, we discuss how the deterministic approach in \citet{Kasy2016} is a case of the \citet{MorganRubin2012} approach, but with the threshold for acceptable randomizations set so high that there is only one acceptable treatment assignment vector---and thus an undefined minimum possible $p$-value and no way to carry out a randomization test. 

This discussion is particularly useful in the context of very small experiments, which can arise in contexts such as clinical medicine and development economics. As \citet{deaton2020} points out, if there is an experiment in which two villages are to be assigned to treatment and two to control, there are only six possible allocations of the treatment conditions, one of which would have been the self-selected allocation. Even with ``hundreds of villages,'' Deaton writes, ``whether or not balance happens depends on how many factors have to be balanced, and nothing stops the actual allocation being the self-selected allocation that we would like to avoid.'' Rerandomization, though, \textit{would} stop the actual allocation from being a specific unbalanced application that we would like to avoid, and we formalize an approach to arriving at an optimum rerandomization that achieves covariate balance without being overly deterministic---thereby allowing use of randomization tests even with close-to-optimal balance. 

The issues discussed here have resonance with long-standing issues raised by Neyman's foundational 1923 paper. As Donald Rubin wrote in 1990, this paper ``represents the first attempt to evaluate, formally or informally, the repeated-sampling properties over their nonnull randomization distributions...this contribution is uniquely and distinctly Neyman's'' (Rubin 1990). The insight that variance could be---and indeed needs to be---calculated in part based on variations over these distributions is a key conceptual piece of the modern potential outcomes framework.

One component of variance calculations over randomization distributions is expressed, in early form, in Neyman (1923)'s description of ``true yields,'' which Neyman describes as `` repeat[ed]...measurement of the yield on
the same fixed plot under the same conditions'' (Neyman 1923). Rubin characterizes this as a way of describing ``[v]ariation in efficacy of randomly chosen
versions of the same treatment'' (Rubin 1990), and this idea---that there can be variation in outcomes over randomly chosen version of the same treatment---that has its echo in current work on rerandomization. Our contribution here is to probe some of the consequences of researcher choices in the rerandomization context for exact significance tests for treatment effects, focusing on the ways in which there may or may not be an optimal degree of randomness in treatment assignment. 


In the remainder of the paper, we answer the question of where experimenters should set the acceptance threshold for rerandomization, especially in small experiments. Our argument proceeds in three parts. First, we discuss how the minimum $p$-value achievable under a rerandomization scheme is a function of the acceptance probability, $p_a$. Then, we discuss how the minimum $p$-value determines the width of fiducial intervals when these intervals are generated using permutation-based tests that account for the rerandomization procedure. Finally, we propose a simple pre-experiment analysis to explore the optimal degree of randomness under varying assumptions about the data-generating process. In brief, the experimenter can use knowledge of the covariates and prior assumptions about the data structure to select the degree of randomness allowed via the rerandomization threshold optimally.

We make these rerandomization tools available in open-source software. The repository, available at
\textbf{\textsf{github.com/cjerzak/fastrerandomize-software}}
contains tutorials on how to deploy this package on real data. 

\begin{table}[H]
\centering
\begin{tabular}{l >{\raggedright\arraybackslash}p{3cm} >{\raggedright\arraybackslash}p{3cm} >{\raggedright\arraybackslash}p{3cm}}
\toprule
Criteria & Neyman & Fisher & Rerandomization \\
\midrule
Objective & Estimate causal effects and obtain unbiased estimates & Randomize to ensure the validity of significance tests & 
Balance covariates to improve causal effect precision \\
\addlinespace[0.9em]
Statistical Assumption & Potential outcomes are fixed; random assignment of $\bW$ & Random assignment of $\bW$ & Random assignment of $\bW$ \\
\addlinespace[0.9em]
Pre-design data requirements &  Knowledge of number of observations (if completely randomizing)  &  Knowledge of number of observations (if completely randomizing)   & Knowledge of number of observations (if completely randomizing); baseline pre-treatment covariates \\
\addlinespace[0.9em]
Efficiency & Focused on deriving efficient estimators & Focused on exactness & Can improve efficiency, maintain exactness \\
%
%
\addlinespace[0.9em]
Balance & Often achieved via stratification & Not guaranteed unless blocking, depends on sample size & Explicitly aims for balance \\
\addlinespace[0.9em]
%
\bottomrule
\end{tabular}
\caption{Comparison of three paradigms in experimental design: Neymanian, Fisherian, and rerandomization-based inference.}
\label{tab:Optimality}
\end{table}

\section{Application to a Modern Agricultural Experiment: An Introduction}\label{S:AppIntro}
We motivate methodological problems discussed in this paper using a recent randomized experiment on agricultural tenancy contracts \citep{burchardi2019}. In this study, researchers randomized the assignment of five different tenancy contracts across three treatment arms to 304 tenant farmers in 237 Ugandan villages. The impetus for this study comes from the longstanding hypothesis in microeconomics that one reason for low agricultural production among tenant farmers is that tenancy contracts which turn a large share (e.g., half) of a tenant farmer's crops to the landowner effectively incentivize low output \citep{marshall1890}. This hypothesis is most prominently due to Alfred Marshall and thus is sometimes called {\it Marshallian inefficiency.} 

To test this empirically, the authors randomized tenants into three treatment arms reflecting three different tenancy contracts: one where tenants kept 50\% of their crop (the standard tenancy contract in Uganda, labeled the Control arm in the study), one where they kept 75\% (labeled Treatment 1 in the study), and one where they kept 50\% of their output but earned an additional fixed payment---either as a cash transfer (labeled Treatment 2A) or as part of a lottery (labeled Treatment 2B). 

Consistent with the idea of Marshallian inefficiency, \citet{burchardi2019} find that, when tenants are able to keep more of their own output (75\% versus 50\%, Treatment 1 and Control in the study), they generate 60\% greater agricultural output. Treatment 2---the additional payment---does not lead to greater output compared to the control tenancy contract.

This experiment has much in common with the experiments that Neyman was researching in the early twentieth century. Its aim is to improve agricultural yields, just as in many of the classic experiments that both Neyman and Fisher designed. And because there is variation between farms, plots, and farmer characteristics, it is critical that, at baseline (pre-experimentation), the studied units are balanced on covariates that could affect farm yield.

This study had a large number of units---large enough that any rerandomization strategy would probably not meaningfully have changed the main results (since rerandomization and t-tests are asymptotically equivalent \citep{proschan2019}). However, it was carried out in partnership with one of the largest non-governmental organizations in Uganda, and researchers unable to work with such a large and well-funded partner might not have the opportunity to carry out such a large trial. Small trials where rerandomization is critical to finding an unbiased treatment effect with a narrow confidence interval are not uncommon in general, especially in clinical medicine \citep{proschan2019}.

\begin{figure}[H]%
    \centering
\subfloat{{\includegraphics[width=.45\linewidth]{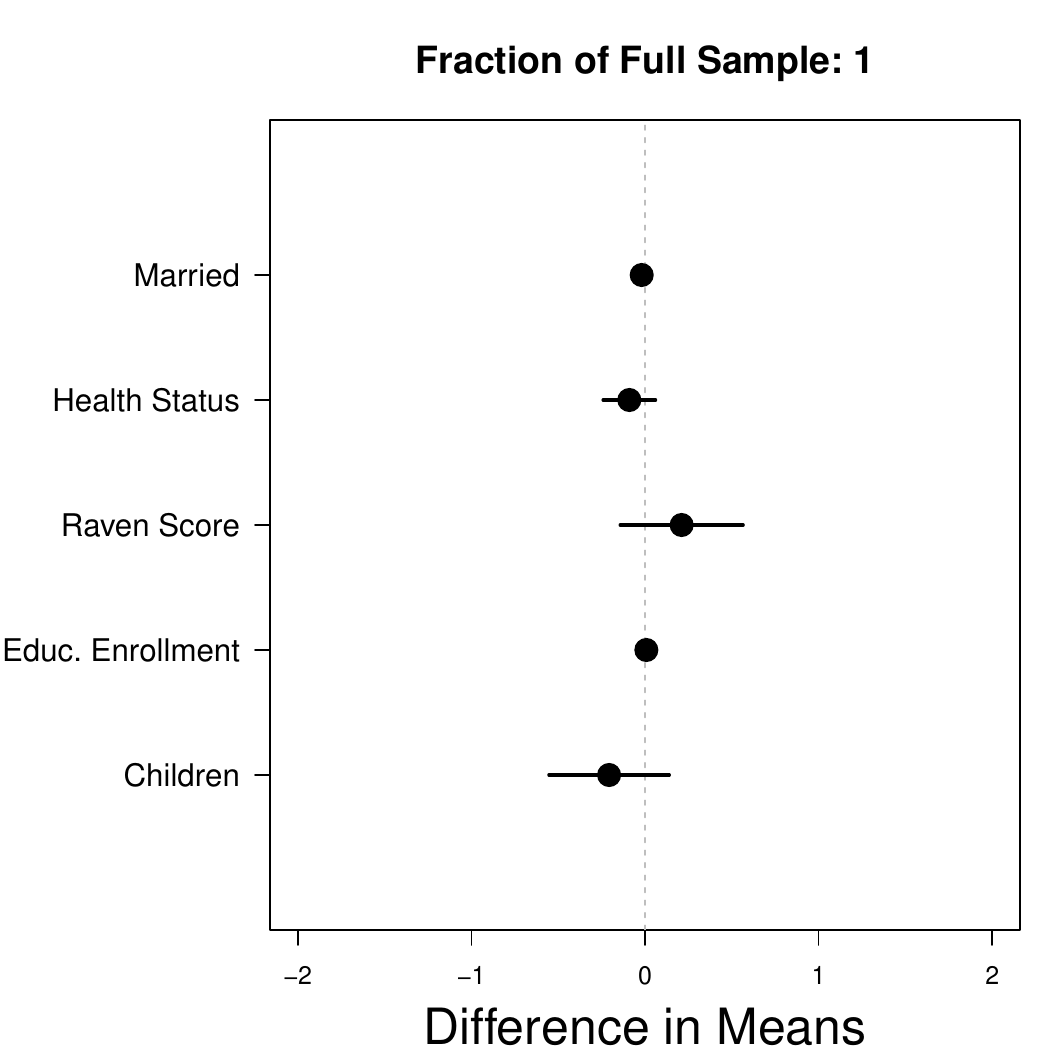} }}%
  \subfloat{{\includegraphics[width=.45\linewidth]{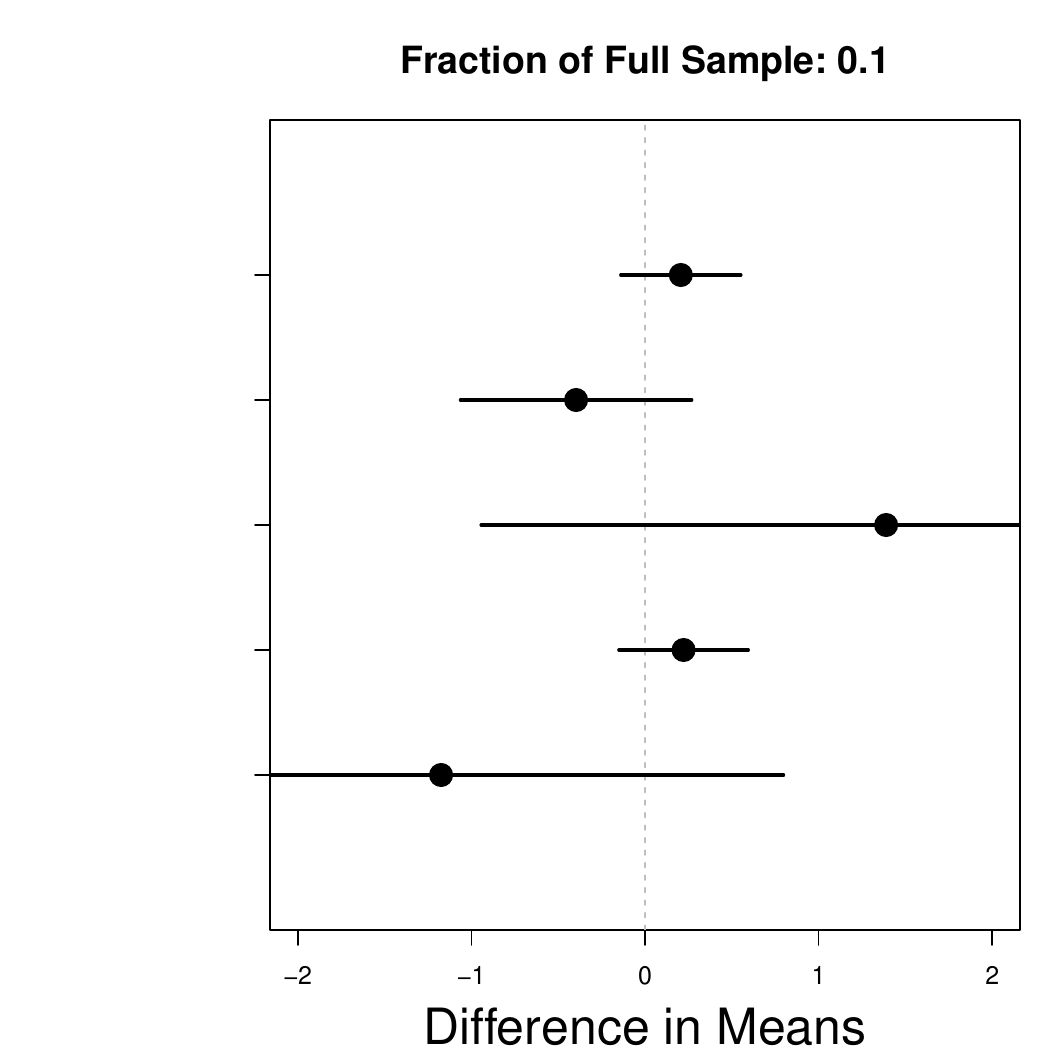} }}%
       \caption{{\it Left panel:} Difference in covariate means in the full data. {\it Right panel:} Difference in covariate means for a randomly selected 10\% sample of the full data.}
   \label{f:PreTreatBal}
\end{figure}

Figure \ref{f:PreTreatBal} displays imbalance in the full and a randomly selected 10\% subset of the Treatment 1 and Control units is shown below. As we see, with a larger sample size, finite-sample balance is approximately maintained, but with a small experiment, there are larger differences between the treatment and control groups. These imbalances  contribute to increased variance in ATE estimates, as an uncertainty estimate for the ATE in the full sample (s.e. $=$ 0.05) is much smaller than in the reduced sample (s.e. $=$ 0.29). Add to this some of the issues associated with small samples regarding degrees of randomness we shall discuss, we see that the study of rerandomization in small studies is of practical importance. 

In this context, could we have improved precision of the treatment effect estimates for a 10\% sample of the data using rerandomization-based approaches? Does rerandomization provide a route to precise treatment effect estimates in the case of possible covariate imbalance? If so, how should the rerandomization process proceed? We turn to this issue in the next section. 

\section{Rerandomization in Small Samples}
\subsection{Notation \& Assumptions}
Before we present our main arguments, it is useful to introduce the notation we will use going forward. Following the notation in \citet{MorganRubin2012}, we assume that the experiment of interest is $2^1$, meaning there is one experimental factor with two levels (i.e., a treatment and control level). Let $W_i = 1$ if unit $i$ is treated and $W_i = 0$ if unit $i$ is control.  Say $i \in \{1, ..., n\}$. The vector ${W}_{n \times 1}$ denotes treatment assignment. Let $\bX_{n \times k}$ denote the data matrix containing $k$ pre-treatment covariates relevant to the outcome of interest. We are interested in the outcome matrix $\mathbf{Y}_{n \times 2}$, where ${Y}(0)_{n \times 1}$ denotes the complete potential outcomes under control and ${Y}(1)_{n \times 1}$ denotes the complete potential outcomes under treatment. If $\left(Y_{\textrm{Obs}}({W})\right)_{n \times 1}$ denotes the vector of observed outcome values, $Y_{\textrm{Obs}, i} = Y_i(1)W_i + Y_i(0)(1-W_i)$. If we assume the sharp null hypothesis (zero treatment effect for every unit) and if we leave ``${Y}_{\textrm{Obs}}$ fixed and simulating many acceptable randomization assignments, ${W}$, we can empirically create the distribution of any estimator,'' $g(\mathbf{X}, {W}, {Y}_{\textrm{Obs}}({W}))$ conditional on the null hypothesis \citep{MorganRubin2012}. We make the additional assumption that the experimenter is interested in using Fisherian inference and that the treatment effects are constant and additive. Next, we define randomization-based fiducial intervals because these intervals are the core of both \citep{MorganRubin2012}'s insight and our contribution to the literature. Researchers can also exploit the duality between intervals and tests to form a fiducial interval from a randomization test. 

\subsection{Rerandomization and the Limits of Randomization Inference}
Is it possible to take rerandomization ``too far'' in the sense that exact tests are no longer informative? We argue that it is indeed possible, in the sense of accepting so few randomizations that it is no longer possible to perform a meaningful randomization test. This important method of assessing significance in exact tests becomes non-informative. 

Consider the following example: there are 8 experimental units, and we would like to test the effect of one factor at two levels (one treatment group and one control group). There is one background covariate, $h$, which denotes the hour a test is conducted. For simplicity, assume that there are 8 observed values for $h$,  ranging from $1$ to $8$. We can quantify the randomization balance with the square root of the quadratic loss: 
\begin{equation}
m = \sqrt{\left[ (\bar{h}|W_i=1) - (\bar{h}|W_i=0)\right]^2}
\end{equation}
where smaller values of $m$ are more desirable because they indicate better balance on $h$. If the experiment is completely randomized and no rerandomization done, the smallest possible $p$-value is $1 \div {8 \choose 4} = 0.015$ and the average $m$ value is $1.4$. If the experiment is pair-matched and complete randomization occurs within pairs for whom $h\in\{1,2\}, ..., h\in\{7,8\}$, the average $m$ improves to $0.375$. However, the minimum possible $p$-value in randomization inference is now $ 1 \div {2 \choose 1}^4 = 0.063$. The minimum possible $p$-value does not vary linearly with the maximum acceptable $m$ in completely randomized experiments (see Table \ref{t:balance_vs_pvalue}). 

\begin{table}[htb]\centering \footnotesize 
\caption{A simple example: How does one's balance threshold influence exact $p$-values?} 
\begin{tabular}{cc} 
\\[-1.8ex]\hline 
\hline \\[-1.8ex] 
Maximum acceptable balance score&Minimum $p$-value \\ 
\hline \\[-1.8ex] 
4& 0.125 \\ 
3& 0.028 \\ 
2& 0.018 \\ 
1& 0.015 \\ 
0& 0.014 \\ 
\hline \\[-1.8ex] 
\end{tabular} %
\label{t:balance_vs_pvalue} 
\end{table}

This example illustrates that low acceptance thresholds can invalidate the estimation of uncertainty in the sense that the minimum possible $p$-value will eventually increase above $0.05$. In addition, the minimum possible $p$-value varies nonlinearly with the level of strictness the researcher adopts in accepting randomizations. These considerations suggest that it may be possible to systematically establish an acceptance criterion that will minimize the minimum possible $p$-value while maximizing the balance improvement from rerandomization.

This insight formalizes the intuition, described in \citet{deaton2020}, that including every possible combination of treatment allocations will inevitably involve including a large number of irrelevant treatment allocations that do not help the researcher understand a possible treatment effect. ``Randomization, after all, is random,'' Deaton writes, ``and searching for solutions at random is inefficient because it considers so many irrelevant possibilities.'' One intuitive way to understand our approach to optimal rerandomization is that it has the advantage of, as Deaton writes, ``provid[ing] the basis for making probabilistic statements about whether or not the difference [between treatment and control groups] arose by chance,'' without the disadvantage of including many irrelevant treatment allocations, including ones which the researcher would specifically like to avoid. 

\subsection{Randomness in Rerandomization and Implications for Exact Randomization Tests}
An important theoretical principle here is that 
\begin{align*}
\textrm{Minimum $p$-value} &= \frac{1}{\textrm{\# acceptable randomizations}}.
\end{align*}
This expression is true because the exact $p$-value is defined as the fraction of hypothetical randomizations showing results as or more extreme than the observed value. If there are $r$ acceptable randomizations, then at least $1$ of the $r$ randomizations is as or more extreme than the observed value, so the minimum fraction is $1/r$. When the set of acceptable randomizations gets smaller, the minimum $p$-value invariably gets larger.

\begin{figure}[htb!]%
    \centering
    \includegraphics[width=.75\linewidth]{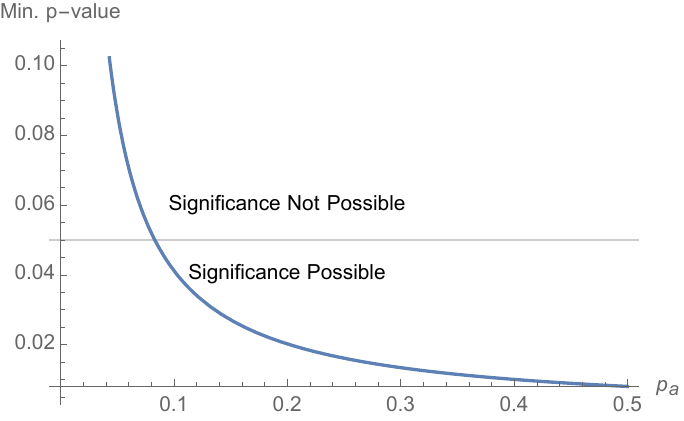}
   \caption{With $n_{\textrm{Cand}}$ held fixed, we can plot the relationship between $p_a$ and the minimum $p$-value.}
   \label{f:analyticalv}
\end{figure}

\begin{figure}[H]%
    \centering
    \includegraphics[width=.55\linewidth]{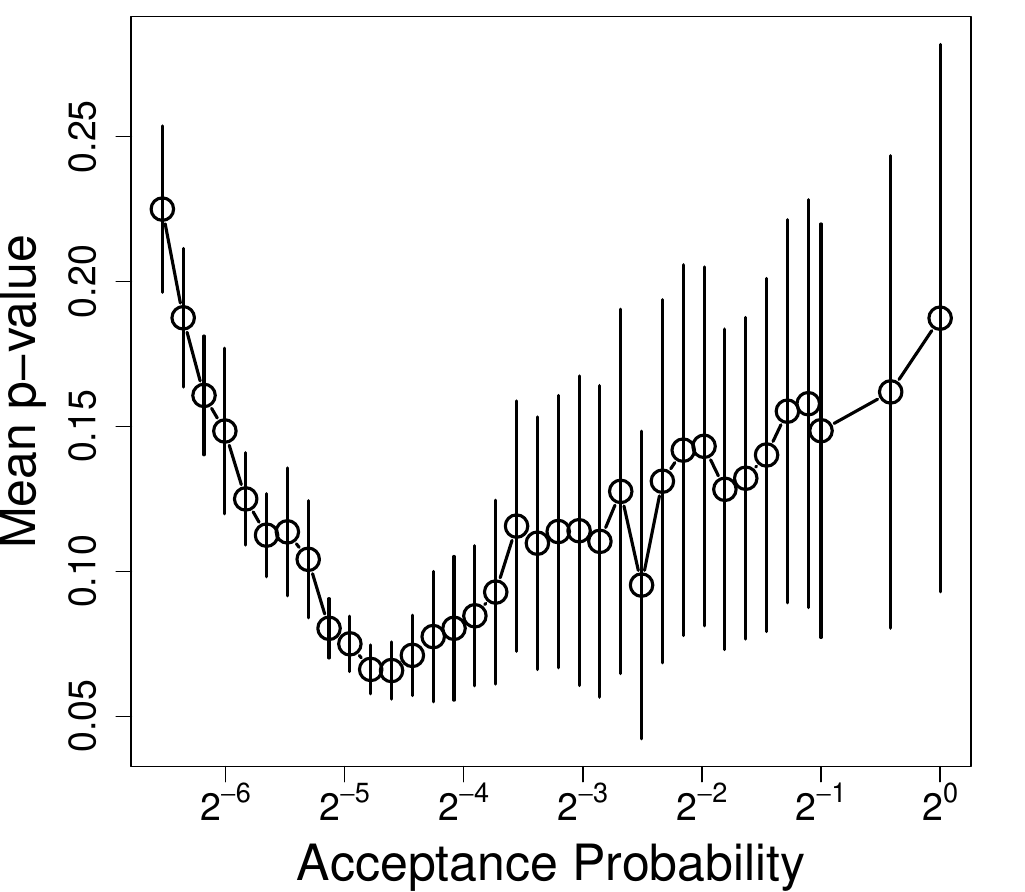}
\subfloat{{\includegraphics[width=.55\linewidth]{RerandomizationViz.pdf} }}%
       \caption{ Illustrating the relationship between acceptance probability and expected $p$-value in small samples. Error bars are 95\% confidence intervals across simulations.  
   \label{f:SimExample}}
\end{figure}

Setting $n_{\textrm{Cand}} = { 10 \choose 5 }$ in Figure \ref{f:analyticalv}, we see the clear non-linear relationship between $p_a$ and the minimum $p$-value. This figure is consistent with the later Monte Carlo results of Figure \ref{f:SimExample}. 

The preceding expression shows how there is a point after which, as experimenters reduce further $p_a$, fiducial intervals get wider as well, since it becomes more and more difficult to reject $H_0:\tau = \tau_0$ for any $\tau_0$. In the extreme, if the minimum $p$-value is above $\alpha$, randomization-based fiducial interval will become non-informative at $(-\infty, \infty)$. 

For the sake of illustration, we briefly examine an example from simulations that we will later explore more carefully (see \S\ref{s:SimDesign} for design information). For each candidate randomization, we calculate $M$, where 
\begin{align*}
M &:= \left( \bar{ \mathbf{X} }_T  - \bar{ \mathbf{X} }_C \right)' \bigg[ \textrm{Cov}\left( \bar{ \mathbf{X} }_T - \bar{ \mathbf{X} }_C \right)^{-1} \bigg] \left( \bar{ \mathbf{X} }_T - \bar{ \mathbf{X} }_C \right); 
\\ &= n p_w \left(1 - p_w \right) \left( \bar{ \mathbf{X} }_T - \bar{ \mathbf{X} }_C \right)'  \textrm{Cov}(\mathbf{x})^{-1} \left(\bar{ \mathbf{X} }_T - \bar{ \mathbf{X} }_C \right), 
\end{align*}
where $p_w$ denotes the fixed proportion of treated units and $\textrm{Cov}(\mathbf{x})$ denotes the sample covariance matrix of $\mathbf{X}$. We then accept the top $p_a$-th fraction of randomizations (in terms of balance) across all possible randomizations.\footnote{Note: We could also accept $M$ with probability $p_a$ based on the inverse CDF of $M$. Due to the Multivariate Normality of $\textbf{X}$, we see that $M\sim \chi_k^2$; results are similar across approach.}


We see in Figure \ref{f:SimExample} how, in small samples, different rerandomization acceptance thresholds induce different expected exact $p$-values, where the expectation averages across the data-generating process. More stringent acceptance probabilities reduce imbalance and therefore finite-sample bias and sampling variability of estimated treatment effects (see \citet*{Li9157}, also \citet{BraMir19}). However, at a certain point, the size of the acceptable rerandomization set no longer supports significant results, as eventually there is only one acceptable randomization (leading to a minimum and expected $p$-value of 1). 

As a consequence of this discussion, as the acceptance threshold gets more stringent, despite improved sampling variability of the treatment effect estimate, exact fiducial intervals will eventually become non-informative as the interval width goes to $\infty$ because the size of the rerandomization set can only yield a $p$-value above 0.05. The exact threshold when non-informativity is reached will vary with the experimental design, but the general pattern is theoretically inevitable: if we shrink the size of the set in which a random choice gets made, we also shrink the reference set for assessing uncertainty. This simulation shows that the interval widths do not necessarily get monotonically smaller until the point of non-informativity but can start to increase well before that point (the exact change point will require future theoretical elucidation). In this context, how can we characterize ``optimal'' ways to rerandomize? What sources of uncertainty affect this optimization? 

\begin{figure}[htb!]%
    \centering
\subfloat{{\includegraphics[width=.45\linewidth]{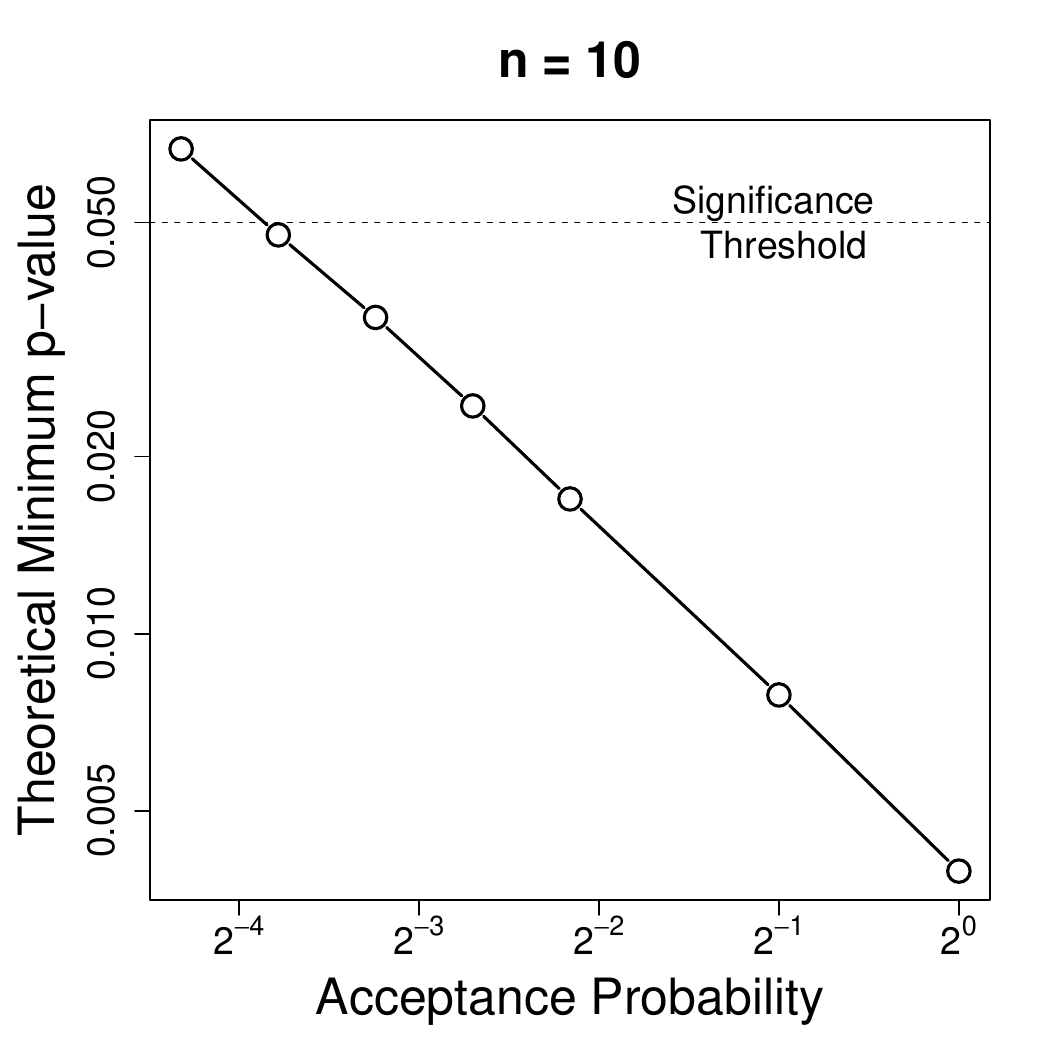} }}%
       \caption{ As the acceptance probability decreases, the minimum $p$-value increases until no statistical significance at $\alpha = 0.05$ is possible.}
   \label{f:figs_minp}
\end{figure} 

\noindent We can state these ideas more formally as follows. 

\vspace{0.5cm} \noindent  Let $\mathcal{R}$ denote the set of all possible ways of assigning $N$ observational units to treatment or control. Let $\mathcal{A}$ denote the set of all algorithms that take $\mathcal{R}$ as an input and give as an output a reduced set of acceptable assignments, $\mathcal{R}_A$. A subset of $\mathcal{A}$, denoted $\mathcal{A}^1$ produces an $\mathcal{R}_A$ such that $|\mathcal{R}_A| = 1$. This subset contains \citet{Kasy2016}'s algorithm as a member, as well as any algorithm for returning a strictly optimal treatment vector (according to some imbalance measure).


\vspace{0.5cm} \noindent \textbf{\TheoremName{} 1}. Any member of the set, $\mathcal{A}^1$, yields infinitely wide fiducial intervals and has no ability to reject any null hypothesis using randomization tests. 

\vspace{0.5cm} \noindent  \textbf{Proof of \TheoremName{} 1}. By construction, the minimum possible $p$-value of non-parametric randomization tests for any member of $\mathcal{A}^1$ is 
\begin{align*}
\textrm{Minimum $p$-value} &= \frac{1}{\textrm{\# acceptable randomizations}}.
\end{align*}
When $|\mathcal{R}_A| = 1$ by definition, we have 
\begin{align*}
\textrm{Minimum $p$-value when $|\mathcal{R}_A|$  is $1$} &= \frac{1}{1} = 1 > 0.05. 
\end{align*}
Thus, all $p$-values will be greater than $0.05$ for $\mathcal{A}^1$, directly implying no ability to reject any exact null hypothesis. Moreover, by the construction of intervals from tests, it follows that every candidate endpoint of the fiducial interval will fail to be rejected. As a consequence, the fiducial interval will span the entire range of the outcome variable and will be infinitely wide for continuous outcomes. In sum, optimal treatment assignment leads to non-informative exact tests and confidence intervals. (We note that the problem of obtaining Neymanian randomization-based  distributions for the test statistic is different and outside the scope of this work; see the discussion in \citet{ImbensRubin2015,branson2018sampling}). 

There is an interesting connection to the discussion of randomness in rerandomization approaches here to the literature on matching contains a non-informativity of a similar nature. In that context, researchers must decide how much dissimilarity to allow between matches. This dissimilarity is often set with a caliper. \citet{CochranRubin1973}, \citet{Austin2011}, and others have made recommendations about the optimal caliper size in the context of propensity score matching.  These recommendations seek to reduce the imbalance as much as possible without ending up with no actual matches: if the caliper were exactly 0 and if the sample space were continuous, the analysis would be non-informative as no units could be matched. In both matching and rerandomization, then, we would like to create a situation where treated and control units are as similar as possible to one another. However, whereas achieving no acceptable matches affects point estimates, achieving a single acceptable randomization affects the evaluation of null hypotheses using exact inference. 

This section illustrates an important tradeoff in rerandomization involving small samples: more stringent acceptance thresholds improve balance and reduce sampling variability in treatment effect estimates, but too stringent thresholds lead to non-informative hypothesis tests. Balancing these considerations, the choice of the rerandomization acceptance probability $p_a$ (or equivalently, the threshold $a$) is critical. In the next section, we will explore several approaches for setting the acceptance threshold. 

\section{Approaches to Choice of Rerandomization Threshold}
The choice of $p_a$ involves considering multiple factors including improved covariate balance, number of acceptance randomizations, and computational time to randomization acceptance. The first two issues have been discussed already. The last issue is relevant because of the waiting time until an acceptable randomization is distributed according to a Geometric($p_a$) distribution in the Multivariate Normal case. In general, we would expect to generate $1 / p_a$ randomizations before accepting a single one. In the context of this tradeoff, there are several possible approaches.  

\subsection{ An \emph{a priori} threshold}
As Morgan and Rubin write, ``for small samples, care should be taken to ensure the number of acceptable randomizations does not become too small, for example, less than $1000$'' (p. 7, \citet{MorganRubin2012}). This view suggests that $p_a$ should be set as low as possible so long as the number of acceptable randomizations is greater than a fixed threshold that is determined \emph{a priori}. In this sense, one decision rule might be to set $p_a$ such that 
\begin{align*}
\textrm{minimum $p$-value given $p_a$}&= \beta. 
\end{align*}
The value of $p_a$ which yields $\beta$ can easily be found by inverting the formula for the minimum $p$-value. On the one hand, it is difficult to know whether a given threshold is reasonable given the number of units available, the number of covariates observed pre-treatment, and the computational resources at researchers' disposal. On the other hand, a threshold is easy to interpret and does not involve additional optimization steps. 

\subsection{Heuristic Tradeoff}\label{s:Heuristic}
Our simulations illustrated how the fiducial interval widths do not get monotonically smaller with $p_a$ until the point of non-informativity; rather, they can start to increase before that point in small samples. Thus, in selecting $p_a$, there is a tradeoff between the variance gains from rerandomization and the decreasing size of the acceptable randomization set. We could therefore propose an alternative procedure to determine a $p_a$ value that explicitly trades off the costs and benefits of rerandomization. 

Following \citep{MorganRubin2012}, let $e_{p_a}(p_a) = e_{a}(a) := \textrm{minimum $p$-value given $p_a$}$. Let $v_{p_a}(p_a) = v_a(a) := v_a$, where $v_a$ loosely denotes the remaining variance between treatment and control covariate means on a percentage basis. This notation allows us to emphasize the duality between setting $a$ and $p_a$, where $p_a$ is defined as the acceptance probability and $a$ is defined as the acceptance threshold for $M$. 

We want both the minimum $p$-value given $p_a$ and $v_a$ to be small: when the minimum $p$-value is small, we can form tight fiducial intervals; when $v_a$ is small, we achieve a large reduction in variance for each covariate. Furthermore, both $e_{p_a}(p_a)$ and $v_{p_a}(p_a)$ are bounded between $0$ and $1$. Thus, we want to set $p_a$ (or, equivalently $a$) such that 
\begin{align*}
p_a^* &:= \textrm{arg}\min_{p_a} \;\;\; \lambda \cdot e_{p_a}(p_a) + (1-\lambda) \cdot v_{p_a}(p_a), 
\end{align*}
where $\lambda \in [0,1]$ determines the tradeoff between the variance reduction and the minimum possible $p$-value. It is possible to derive an efficient solution to this optimization problem using gradient descent.

The optimization framework described here has several attractive features. First, given a minimum $p$-value, we can also back out the $\lambda$ value that this implies. Thus, this framing therefore allows for the evaluation of threshold choices. Second, when $w = 0.5$ and there is only one covariate, the minimum $p$-value will always be less than 0.05. Third, conditional on number of covariates, $p_a^*$ describes the same relative place on the objective function no matter the sample size, so in this sense, it is comparable across experiments. Finally, $p_a^*$ has an intuitive interpretation in that it captures the point at which we have gained most of the variance-reducing benefits of rerandomization but have not incurred significant inferential costs. 


\subsection{Optimal Rerandomization Threshold with Prior Design Information}\label{s:PriorDesignIntro}
The preceding discussion of heuristic tradeoffs describes dynamics regarding the choice of randomization threshold when no prior information is available. However, when prior design information on units is accessible, we can incorporate that information explicitly to help improve exact tests in a design-based manner.

Considerations of optimal precision have a long history in the analysis of treatment effects. For example,  \citet{neyman1933zarys} and \citet{neyman1952recognition} discuss minimizing confidence interval width via stratum-size informed stratified sampling \citep{kubiak2022prior}. In general, in the Neymanian framework, optimal experimental design is that which generates unbiased and minimal variance estimates, usually under the condition of no prior assumptions on the data-generating process. 

While discussions of power in the rerandomization framework are developed in \citet{branson2022power} and the weighting of covariates in the rerandomization \citep{liu2023bayesian}, there is less guidance in the literature about how to consider optimality in randomization choice from a Neymanian-informed Fisherian perspective---a task we turn to in this section. 

Because rerandomization-based inference is simulation-based, the {\it a priori} calculation of the optimal acceptable balance threshold is difficult (although as mentioned  asymptotic results are available \citep{Li9157}). We  here show how we can use prior information on background covariates and (b) prior information about the plausible range of treatment effects to select the covariate balance threshold so as to explicitly minimize the acceptable randomization threshold minimizing the expected $p$-value. The expectation is taken over the aforementioned sources of prior information, with the intuition being that when we have an estimate for how prior information is relevant for predicting the outcome, we can use this estimate for determining how to select an optimal point on the tradeoff between better balance and sufficient randomness to perform exact rerandomization inference. 

Formally, we select the rerandomization threshold using knowledge of the design---in particular, the covariates and prior assumptions on plausible distributions over treatment effect and relationships between the baseline covariates and outcome: 
\begin{align*} 
p_a^* = \arg\min_{p_a} \;\; \E_{\bT\sim \bD(p_a),\beta,\btau} \left[ \textrm{$p$-value}(\widetilde{\bY}(\bX, \bT, \btau), \bT)\right],
\end{align*} 
with $\btau$ being the vector of individual treatment effects, $\beta$ defining the parameters of the potential outcome model, denoted using $\widetilde{\bY}$, and $\bD(p_a)$ defining the distribution over the treatment assignment vector as a function of the acceptance probability, $p_a$. The approach here involves simulating potential outcomes under prior knowledge, calculating the expected $p$-value under various acceptance thresholds, and selecting the threshold minimizing this quantity (thereby maximizing power to reject the exact null).

By deciding on the acceptance threshold before analyzing the data, we use the principles of design-based inference pioneered by Neyman and contemporaries. And, by incorporating rerandomization, we improve treatment effect estimates while maintaining the ability to calculate valid exact intervals in a way that incorporates information about the structure of randomness in the randomization inference. 

In summary, selecting the best rerandomization threshold in a given experiment is difficult to do in small experiments given without guidance from asymptotic results. However, we can use simulation-based methods to form an approximation for how $p$-values will on average operate under the prior design knowledge. These approximations are then useful in pre-analysis design decisions, especially as investigators can select the acceptance threshold minimizing {\it a priori} expected $p$-values. 

\section{Exploring Degrees of Randomness in Rerandomization via Simulation}
\subsection{Simulation Design}\label{s:SimDesign}
We explore the impact of varying levels of randomness in rerandomization on inference through simulation. We simulate potential outcomes under a linear model: 
\begin{align*}
Y_i(0) &= \bX_i' \bbeta + \epsilon_i^{Y(0)}
\\ Y_i(1) &= \tau + \bX_i' \bbeta + \epsilon_i^{Y(1)},
\end{align*}
where covariates, $\bX_i$ are drawn from a Multivariate Gaussian with diagonal covariance and where $\epsilon_i^{Y(t)}$ are independent Gaussian error terms. We study an environment where errors are $N(0,0.1)$, allowing us to assess robustness to noise in the potential outcomes. We also study results across low and high treatment effect strength conditions (where $\tau \in \{0.1,1\}$). Finally, we vary the number of observations, $n\in \{6, 12, 18\}$. 

We then generate a sequence of acceptable randomizations, ranging from 10 to the set of all possible completely randomized treatment vectors, to evaluate performance at different rerandomization acceptance thresholds. We run the pre-analysis design procedure described in \S\ref{s:PriorDesignIntro}, which returns an estimated expected $p$-value using the prior design information. We compute the estimated expected $p$-value from each Monte Carlo iteration. We then compare the range of estimates against the true expected $p$-value minimizer, analyzing both bias and relative RMSE, where for interpretability the RMSE estimate has been normalized by the expected RMSE under uniform selection of the acceptable randomization threshold, $p_a$. We assess performance averaging over randomness in the covariates and outcome. 

\subsection{Simulation Results}\label{s:SimResults}
\begin{figure}[H]%
    \centering
\subfloat{{\includegraphics[width=.95\linewidth]{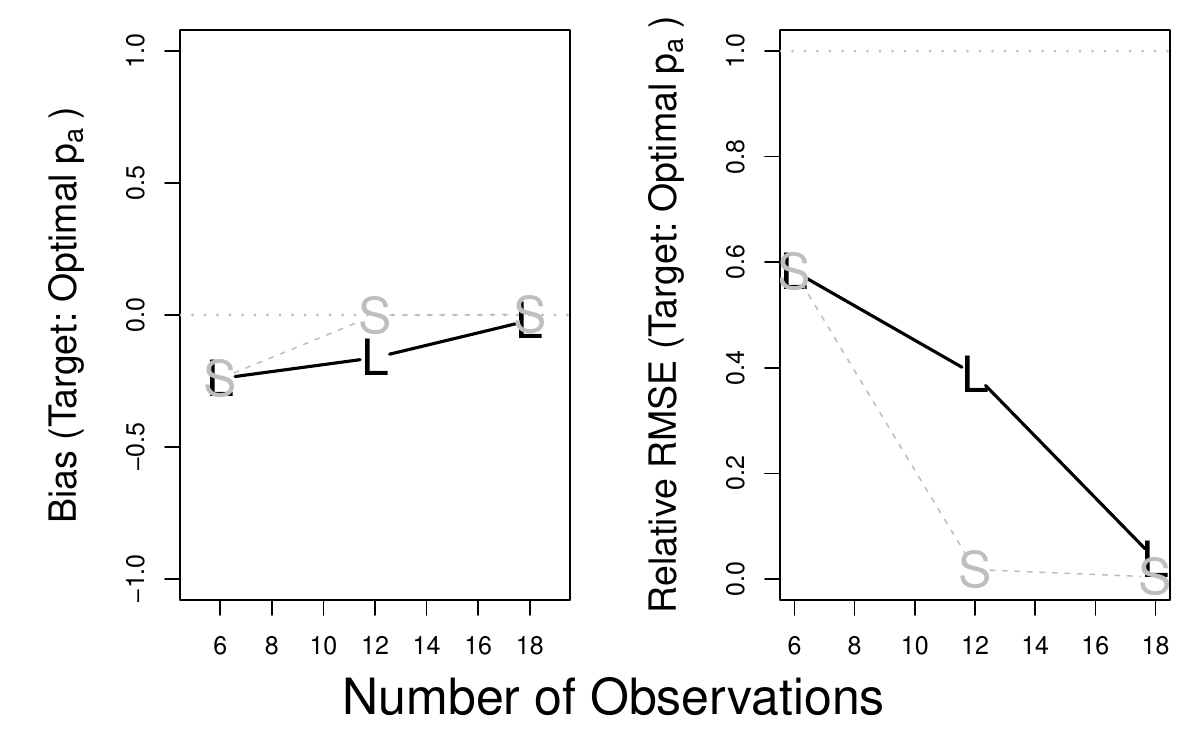} }}%
       \caption{{\it Left panel:} Bias, where the target estimand is the optimal $p_a$ in minimizing expected $p$-value. ``S'' denotes the small treatment effect size case; ``L'' denotes the large treatment effect size case. {\it Right panel:} Relative RMSE, with raw RMSEs have been made relative using the RMSE under uniform selection of $p_a$ as a baseline.}
   \label{f:SimResults}
\end{figure}

Figure \ref{f:SimResults} displays the main results. In the left panel, we see that the Bayesian selection procedure for the rerandomization threshold described in \S\ref{s:PriorDesignIntro} is somewhat downwardly biased under the non-informative priors specified here, both in the small (``S'') and large (``L'') treatment effect cases. In other words, it yields an acceptance probability that is somewhat lower than what we find to be optimal via Monte Carlo. However, in the right panel, we see that the procedure generates lower relative RMSE compared to uniform selection of $p_a$ across the interval $[0,1]$. This finding, again robust to treatment effect size, indicates that there is a systematic structure in the randomness surrounding the rerandomization problem that can be leveraged in pre-design analysis decisions. 

Overall, we find a large decrease in relative root mean squared error (RMSE) when using the pre-analysis rerandomization threshold selection procedure described \S\ref{s:PriorDesignIntro}. Rerandomization allows robust causal inference in cases with high noise levels and smaller sample sizes. 

\section{Application to a Modern Agricultural Experiment, Revisited}

Having shown the possible improvements with our optimal rerandomization approach via simulation, we now explore the same dynamics using data from the real agricultural experiment on tenancy contracts discussed in \S\ref{S:AppIntro}. 

To do this, we first take a random 10\% subsample of the units in the experiment, to illustrate what it would have been like to carry out a much smaller, less costly experiment than done by the original authors. We then carry out a semi-synthetic simulation involving the experimental data. 

 We first fit an OLS model on the full data to estimate outcome model parameters (along with variance-covariances). We then sample these coefficients from a Multivariate Gaussian to simulate counterfactual outcomes for the potential outcomes of each unit in random 10\% subsamples. We need this semi-synthetic simulation protocol to generate the complete potential outcomes table. We use the fixed covariate profiles of units, and average uncertainty in the outcome imputations when reporting assessing performance. 

\begin{figure}[H]%
    \centering
  \includegraphics[width=.45\linewidth]{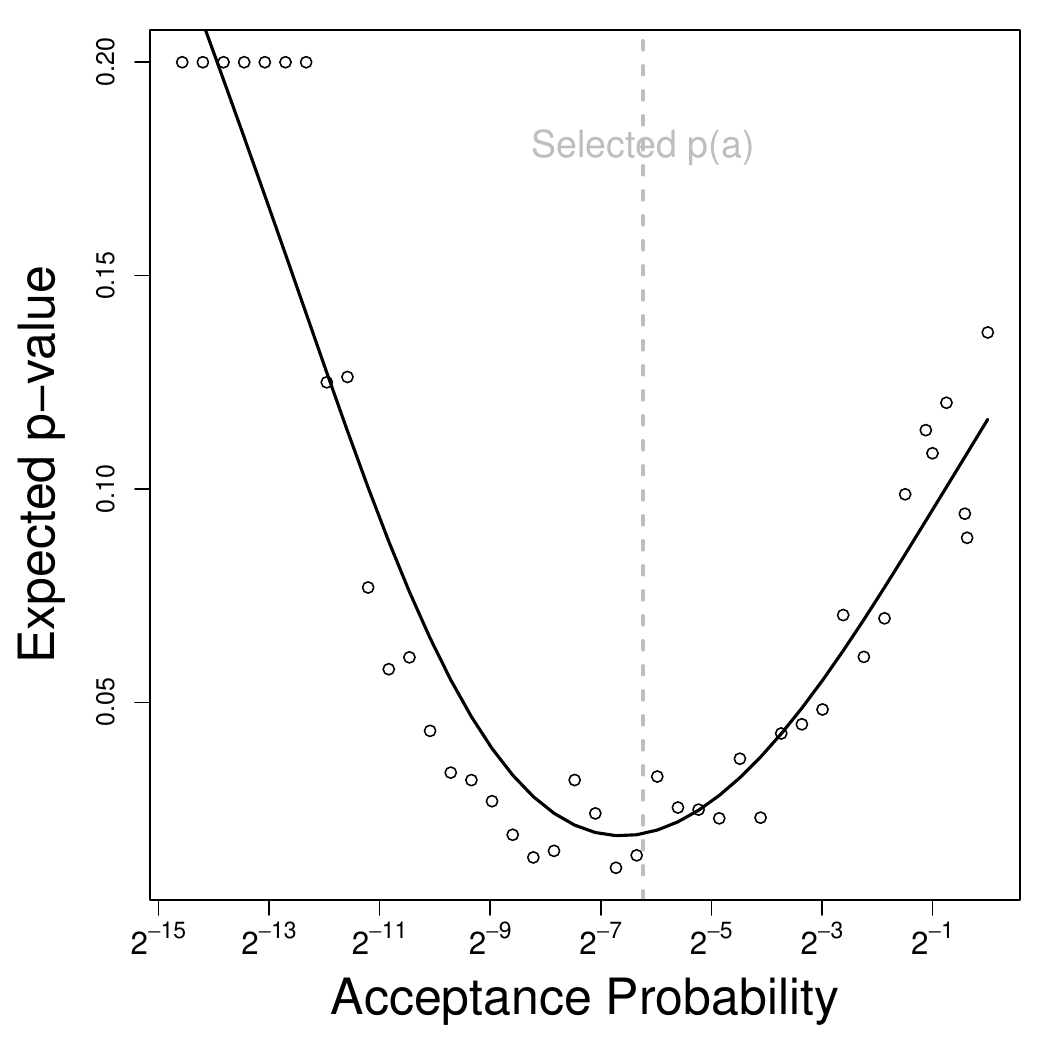}%
       \caption{The acceptance probability minimizing the expected $p$-value is quite close to the one selected by the choice procedure described in \S\ref{s:PriorDesignIntro}.}
   \label{f:AppFig}
\end{figure}

With this semi-synthetic setup where true causal effects are known (and non-zero), we find, as shown in Figure \ref{f:AppFig}, that the true acceptance probability that minimizes the expected $p$-value is 0.008. This is quite close to the selected $p_a$ value from the pre-design procedure described in \S\ref{s:PriorDesignIntro} (value near 0.008; non-informative Gaussian priors for assumed structural parameters used). The implication of these results for practice are that investigators should in this case employ some rerandomization so that there are about 128 acceptable randomization in terms of balance out of the set of all possible 184,756 complete randomizations. When they do so, they would based on the prior design structure expect a $p$-value of 0.027 compared to 0.10 without any rerandomization. 

\begin{figure}[htb!]%
    \centering
  \includegraphics[width=.45\linewidth]{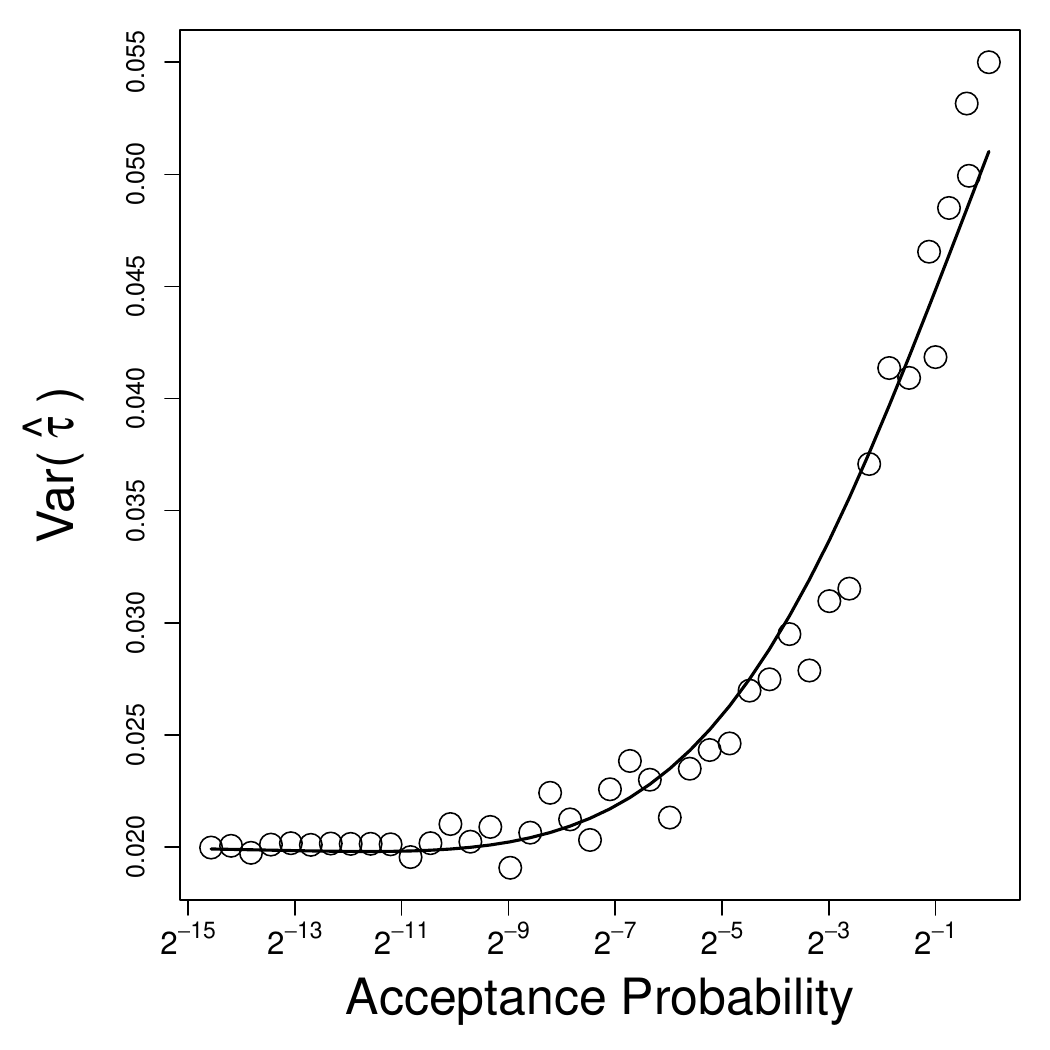}%
       \caption{As we decrease the acceptance probability, the expected $\hat{\tau}$ gets closer to the true value (indicated by the gray horizontal line). The acceptance probability indicated by the aforementioned procedure weights the expected precision gains in $\hat{\tau}$ vs. the uncertainty estimates in the exact $p$-values.}
   \label{f:AppFig2}
\end{figure}

We next observe in Figure \ref{f:AppFig2} the degree to which the difference-in-means estimates for the ATE fluctuate across repeated samples in relation to the acceptance probability. When the acceptance probability remains high, there's a corresponding increase in the average imbalance---and therefore sampling variability. As the acceptance probability diminishes, estimation variance shrinks in tandem with improved balance. The selection procedure elucidated previously selects an acceptance probability that not only ensures a precisely estimated ATE (where precision is meant in terms of sampling variability) but also a testing framework leading to improved power to reject the null, evident from the minimum expected $p$-value we noticed in Figure \ref{f:AppFig}.

Without rerandomization, the finite-sample $\hat{\tau}$'s deviate 14\% or more from the true $\tau$ in this semi-synthetic design, whereas with the rerandomization specified by the pre-experimental design analysis, they deviate by less than 5\%. 

\section{Discussion of Limitations \& Future Work}\label{s:Assumptions}

The discussion presented in this paper relies on several key assumptions that are worth highlighting.

First, we focus our discussion on Fisherian permutation-based inference. As mentioned, the story of how significance tests interact with rerandomization threshold choice may differ for Neymanian repeated sampling inference. However, we note that some of the principles we have discussed here apply (at least in the extreme) for other kinds of randomization-based inference that test null hypotheses beyond the exact null---e.g., the conditional independence null used in \citet{candes2018panning} where randomization inference is performed using draws from the assignment distribution over treatment vector. 

Second, our simulation results assume additivity and constancy of treatment effects. This rules out treatment effect heterogeneity or interactions between covariates and treatment. While rerandomization can still be beneficial under weaker assumptions, the relationship between the acceptance probability and inference may differ in more complex settings. Investigating rerandomization with heterogeneous effects is an important area for continued work.

Finally, we note that the task of achieving very low rerandomization thresholds is computationally intensive for experiments involving many units. While the computational tools introduced in the \verb|fastrerandomize| package accompanying paper make possible the analysis of the full permutation set over treatment vectors ranging in size into the hundreds of millions using an accelerated linear algebra compiler, further computational effort to adaptively build this set would enable researchers to achieve better balance while maintaining use of exact inference or conditional randomization tests. 

\section{Conclusion}
In this paper, we formalize the idea that rerandomization can be taken ``too far,'' in the sense that the variance-reduction benefits of rerandomization can be outweighed by the costs of having undesirably wide permutation-based fiducial intervals when acceptance probabilities are low. We accomplish this formalization by deriving the minimum possible $p$-value, which turns out to be a function of the acceptance threshold and the number of units. 

In this light, we presented a unified approach to the problem of determining an optimal acceptance threshold. While the decision for a rerandomization threshold traditionally operates in the absence of prior information, the availability of prior design details helps structure the randomness in a way that can be leveraged in the design phase. By leveraging this information, researchers can hone experimental precision in a design-centric fashion, a concept rooted in the foundations set in \citet{neyman1933zarys} and \citet{neyman1952recognition}. Drawing from the Neymanian ethos, the ideal experimental design offers unbiased and variance-minimized estimates, typically devoid of data-generating presumptions. Notably, as the rerandomization framework is inherently simulation-based, determining an \textit{a priori} optimal balance threshold can be intricate, despite some extant asymptotic results \citep{Li9157}. 

Although Fisherian and Neymanian approaches to experimental treatment effect estimation differ, and although rerandomization---because it is grounded in randomization inference---is more aligned with a Fisherian approach, both researchers were concerned with the efficient use of data. Rerandomization can make better use of available data by improving covariate balance, thereby serving the interests of efficiency from both Fisherian and Neymanian perspectives.

We have contributed to the understanding of rerandomization as a kind of slippery slope: at the top of this slope, all randomizations are accepted and, at the bottom, only one randomization is accepted but it becomes impossible to estimate uncertainty using important non-parametric tests. Our paper thus helps experimenters to understand where they are on this slope, where they might most prefer to be, and how they can arrive there. 

Over thirty years ago, \citet{rubin1990} pointed out that, even though Fisher's and Neyman's approaches to hypothesis testing were distinct, they were ``complementary.'' While Fisher relied on a sharp null and Neyman took a repeated-sampling approach over nonnull distributions, both have the intuition that examining all ways that treatments could be assigned can help researchers understand how the treatment changed outcomes for treated units relative to similarly situated control units. In this paper, we hope to suggest one method that uses these same intuitions to make a methodological contribution in today's very different world of ubiquitous randomized experiments and immense computing power. By strategically winnowing down the set of possible randomizations to an optimum number, we contribute an approach that emphasizes balance while reaping the benefits of random variation. 
 \vspace{\baselineskip}  \\

\noindent \textbf{\textsf{Data availability statement}}: Data and code to replicate results, as well as a tutorial in installing and using the \textbf{\textsf{fastrerandomize}} package, are or will be available at \textbf{\textsf{github.com/cjerzak/fastrerandomize-software}}.

\printbibliography


\newpage 
\section{Appendix} 
\subsection{Defining Randomization-based Fiducial Intervals}
Randomization intervals incorporate randomness only through variation in the treatment vector, and do not explicitly make a reference to repeated experiments (as in Neyman's classical confidence intervals). In this case, we can produce an interval by finding the set of all null hypotheses that the observed data would fail to reject. If we recall the constant and additive treatment effect assumption (${Y}(1) = {Y}(0) + \tau$), then the hypotheses can be written as 
\begin{align*}
H_0 &: \tau = \tau_0 
\\ H_1 &: \tau \ne \tau_0. 
\end{align*}
An $\alpha$-level fiducial interval for $\tau$ consists of the set of $\tau_0$ such that the observed test statistic would not lead to a rejection of the null hypothesis at significance-level $\alpha$. Here, when $\tau_0  \ne 0$, the randomization test is conducted by constructing 
\begin{equation*}
Y_i(0)^* = \left( Y_{\textrm{Obs}, i}  - \tau_0 \right) W_i + Y_{\textrm{Obs}, i} ( 1 - W_i).
\end{equation*}
 Then, keeping $Y_i(0)^*$ fixed, we permute the treatment assignment vector, and calculate ${Y}_{\textrm{Obs}}^*$ by adding $\tau_0$ to the treatment group outcomes under the permutation. This procedure generates a distribution of $\hat{\tau}$ under the null $H_0: \tau = \tau_0$. We can use this distribution to calculate a $p$-value for $\hat{\tau}_{\textrm{Obs}}$ under the null.
We can form an $100 ( 1 - \alpha) \%$ interval for $\tau$ by finding the values of $\tau_0$ which generate a $p$-value greater than or equal to $\alpha$. \citet{Garthwaite1996} discuss an efficient algorithm for obtained randomization-based fiducial intervals, which searches for the interval endpoints using a procedure based on the Robbins-Monro search process. 

\subsection{Proofs of \TheoremName{s} 1}
In this section, we define $V:=W^C$ for notational convenience. We assume that the covariates are Multivariate Gaussian and the imbalance metric is the Mahalanobis distance between treated and control means. Then, we can appeal to the finite-population Central Limit Theorem to assume $\mathbf{\overline{X}}_T - \mathbf{\overline{X}}_C$ is Multivariate Normal and therefore that $M(V) \sim \chi_k^2$ \citep{DinLi17}. 

\subsubsection{Proof of \TheoremName{} 1}
For rerandomization, each accepted $V$ is equally likely to have been drawn (since each $V$, accepted or not, was drawn with probability $1 / n_{\textrm{Cand}}$). Recall that, if all outcomes are equally likely, the probability of the event $\{V = v\}$ given $M(V) \leq  a$ is equal to the number of outcomes in which that event occurs (which is $1$) divided by the total number of outcomes (which is $|\mathcal{R}_A|$). This means 
\[
\Pr(V = v|  M(V) \leq a ) = \frac{1}{|\mathcal{R}_A|}. 
\]
Another way to see this is to consider how 
\begin{align*} 
\Pr(V = v| I\{ M(V) \leq a \} ) &= \Pr(V = v|  M(V) \leq a ) 
\\ &= \frac{\Pr( I\{ M(V) \leq a\} |V = v) \times 
\Pr(V = v)}{ \Pr(I\{M(V) \leq a\})}; 
\\ &= \frac{I\{M(v) \leq a\} \times 2^{-k}}{ \frac{|\mathcal{R}_A|}{n_{\textrm{Cand}}}} = \frac{1 \times \frac{1}{n_{\textrm{Cand}}}}{ \frac{|\mathcal{R}_A|}{n_{\textrm{Cand}}}} = 1 / |\mathcal{R}_A|, 
\end{align*} 
where the last line uses the fact that $n_{\textrm{Cand}} = 2^{k}$. The average waiting time for the first acceptance is governed by a Geometric distribution with probability parameter $p_a$. By the independence of the sampling process, the waiting time until the $k^{\textrm{th}}$ acceptance is given by $k / p_a$. 

\subsection{Kasy (2016) in the Rerandomization Framework} 
We argued in the above that Kasy's optimal assignment procedure is a special case of rerandomization in which the only acceptable randomization is the one that provides optimal covariate balance. Let the $\mathbf{x}$ matrix include all observed covariates that the experimenter seeks to balance. Let $\mathbf{W}$ denote the $n$-dimensional treatment assignment vector indicating the treatment group for each unit. In the notation of Morgan and Rubin, the rerandomization criterion can be written as
\begin{equation*}
\varphi(\mathbf{x, W}) = 
\begin{cases}
1,  &\textrm{if $\mathbf{W}$ is an acceptable randomization}, 
\\ 0, &\textrm{if $\mathbf{W}$ is not an acceptable randomization}. 
\end{cases}
\end{equation*}
With the appropriate $\varphi(\cdot, \cdot)$, Kasy's non-deterministic assignment procedure can be considered a special case of rerandomization. That is, we will obtain Kasy's optimal assignment if we define $\varphi(\cdot, \cdot)$ as follows, letting $R(\cdot)$ denote either Bayesian or minimax risk, letting $\hat{\beta}$ denote the choice of estimator, and letting $U$ denote a generic randomization procedure independent of $\mathbf{x}$ and $\mathbf{Y}$, 
\begin{equation*}
\varphi(\mathbf{x, W}) = 
\begin{cases}
1,  &\textrm{if $\mathbf{W}$ minimizes $R(\mathbf{W}, \hat{\beta} | \mathbf{x}, U)$}, 
\\ 0, &\textrm{if $\mathbf{W}$ does not minimize $R(\mathbf{W}, \hat{\beta} | \mathbf{x}, U)$}. 
\end{cases}
\end{equation*}
With $n$ units, we will obtain a $\mathbf{W}$ such that $\varphi(\mathbf{x}, \mathbf{W})=1$ with probability $(1/2)^n$. The expected wait time until obtaining this $\mathbf{W}$ is given by a Geometric distribution, with mean $2^n$. This calculation implies that, as soon as $n\geq 20$, the expected wait-time will exceed $1$ million rerandomization draws if rerandomization is used \citep{SoaresWu1985}. 

This section has put our discussion in closer dialogue Kasy's work in order to further illustrate how Kasy's deterministic procedure for optimal treatment assignment is a special case of rerandomization developed in \citet{MorganRubin2012}. 

\end{document}